\documentstyle[prd,aps,epsf,floats,amsfonts,amssymb,amsmath]{revtex}

\begin{document}
\title{Baryogenesis in $f(R)$-Theories of Gravity}
\author{G. Lambiase and G. Scarpetta}
\address{Dipartimento di Fisica "E.R. Caianiello"
 Universit\'a di Salerno, 84081 Baronissi (Sa), Italy,}
 \address{INFN - Gruppo Collegato di Salerno, Italy.}

\maketitle
\begin{abstract}
$f(R)$-theories of gravity are reviewed in the context of the so
called gravitational baryogenesis. The latter is a mechanism for
generating the baryon asymmetry in the Universe, and relies on the
coupling between the Ricci scalar curvature $R$ and the baryon
current. Gravity Lagrangians of the form ${\cal L}(R)\sim R^n$,
where $n$ differs from 1 (the case of the General Relativity) only
for tiny deviations of a few percent, are consistent with the
current bounds on the observed baryon asymmetry.

\end{abstract}
\pacs{04.50.+h, 98.80Cq, 98.80Es}

The discovery of the accelerated expansion of the Universe
\cite{accUn} has motivated the developments of many models of
gravity. These models are built up, typically, either in the
framework of the conventional General Relativity or in the
framework of its possible generalizations or modifications. In the
last years, among the different approaches proposed to generalize
Einstein's General Relativity, the so called $f(R)$-theories of
gravity received a growing attention. The reason relies on the
fact that they allow to explain, via gravitational dynamics, the
observed accelerating phase of the Universe, without invoking
exotic matter as sources of dark matter. The gravity Lagrangian
for these theories is a generic function of the Ricci scalar
curvature $R$, and the corresponding action with the inclusion of
matter reads
\begin{equation}\label{Lagr}
  S=\frac{1}{2\kappa^2}\int d^4x \sqrt{-g}\, f(R)+S_m[g_{\mu\nu},
  \psi_m]\,,
\end{equation}
where $\kappa^2=8\pi G$. It is a difficult ask to deal with higher
order terms in the scalar curvature, thus the forms of $f(R)$ that
have been most widely studied in literature are $f(R)\sim
R+\epsilon R^m$ and $f(R)\sim R^n$
\cite{capozziello,carroll2004,altri,vollick,barraco}. A careful
and relevant study about the possible form for $f(R)$-Lagrangians
has been performed by Olmo in \cite{olmo}. There it has been shown
that solar system experiments provide strong constraints on the
possible dependence of $f(R)$:
\begin{equation}\label{f-constraint}
  -2\Lambda \leq f(R) \leq R-2\Lambda +l^2 R^2/2\,,
\end{equation}
i.e. the gravity Lagrangian must be {\it nearly} linear in $R$,
with non linear terms bounded by quadratic terms. Therefore the
result (\ref{f-constraint}) excludes non linear terms growing at
low $R$ \cite{carroll1}. In (\ref{f-constraint}) $\Lambda$ is the
cosmological constant.

The aim of this paper is to show that $f(R)$-theories of gravity
provide a framework in which the {\it gravitational} baryogenesis
may occur, and yields the observed baryon asymmetry in the
Universe. The origin of the baryon number asymmetry is still an
open problem of the particle physics and cosmology. Big-Bang
Nucleosynthesis (BBN) \cite{burles} and measurements of CMB
combined with the large structure of the Universe \cite{bennet}
indicate that matter in the Universe is dominant over antimatter;
the order of magnitude of such an asymmetry is
 \[
\eta\equiv \frac{n_B-n_{\bar B}}{s}\lesssim 9\,\, 10^{-11}\,,
 \]
where $n_B$ ($n_{\bar B}$) is the baryon (antibaryon) number
density, and $s$ the entropy of the Universe. As argued by
Sakharov, the baryon asymmetry may be (dynamically) generated if
the following conditions are satisfied \cite{sakharov}: 1)
processes that violate baryon number; 2) $C$ and $CP$ violation;
3) out of the equilibrium. The last point is important because to
have different numbers of baryons and anti-baryons is different,
the reaction should freeze before particles and antiparticles
achieve the thermodynamical equilibrium. However, as shown in
\cite{cohen}, a dynamically violation of CPT may give rise to the
baryon number asymmetry also in a regime of thermal equilibrium.

Recently, within Supergravity theories \cite{kugo}, Davoudiasl et
al. \cite{steinhardt} have proposed a mechanism for generating the
baryon number asymmetry during the expansion of the Universe by
means of a dynamical breaking of CPT (and CP). In this approach
the thermal equilibrium is preserved. The interaction responsible
for CPT violation is given by a coupling between the derivative of
the Ricci scalar curvature $R$ and the baryon current $J^\mu$
\cite{noteJ}
\begin{equation}\label{riccicoupling}
  \frac{1}{M_*^2}\int d^4x \sqrt{-g}J^\mu\partial_\mu R\,,
\end{equation}
where $M_*$ is the cutoff scale characterizing the effective
theory. For different approaches also based on gravitational
baryogenesis, see \cite{lambiase}.

If there exist interactions that violate the baryon number $B$ in
thermal equilibrium, then a net baryon asymmetry can be generated
and gets frozen-in below the decoupling temperature $T_D$. From
(\ref{riccicoupling}) it follows $M_*^{-2} (\partial_\mu
R)J^\mu=M_*^{-2} {\dot R}(n_B-n_{\bar B})$. Therefore the
effective chemical potential for baryons and antibaryons is
$\mu_B={\dot R}/M_*^2 =-\mu_{\bar B}$, and the net baryon number
density at the equilibrium turns out to be (as $T\gg m_B$, where
$m_B$ is the baryon mass) $n_B=g_b\mu_B T^2/6$. $g_b\sim {\cal
O}(1)$ is the number of intrinsic degrees of freedom of baryons.
The baryon number to entropy ratio is \cite{steinhardt}
\begin{equation}\label{nB/s}
  \frac{n_B}{s}\simeq -\frac{15g_g}{4\pi^2g_*}\frac{\dot R}{M_*^2
  T}\Big|_{T_D}\,,
\end{equation}
where $s=2\pi^2g_{*s}T^3/45$, and $g_{*s}$ counts the total
degrees of freedom for particles that contribute to the entropy of
the Universe. $g_{*s}$ takes values very close to the total
degrees of freedom of effective massless particles $g_*$, i.e.
$g_{*s}\simeq g_*\sim 106$. $n_B/s$ is different from zero
provided that the time derivative of the Ricci scalar is
nonvanishing. In the context of General Relativity, the Ricci
scalar and the trace $T_g$ of the energy-momentum tensor of matter
($T_g^{\mu\nu}$) are related as follows
 \[
R=-8\pi GT_g =-8\pi G(1-3w)\rho\,,
 \]
where $\rho$ is the matter density, $p$ the pressure, $w=p/\rho$,
and $T_g=T_{g\, \mu}^\mu$. ${\dot R}$ is zero in the radiation
dominated epoch of the standard Friedman-Robertson-Walker (FRW)
cosmology, characterized (in the limit of exact conformal
invariance) by $w=1/3$. This way no net baryon number asymmetry
can be generated. However, as we shall see, a net baryon asymmetry
may be generated during the radiation dominated era in
$f(R)$-theories of gravity (see \cite{steinhardt,li} for other
scenarios)

The variation of the action (\ref{Lagr}) with respect to the
metric yields the field equations
\begin{equation}\label{fieldeqs}
  f' R_{\mu\nu}-\frac{f}{2}\, g_{\mu\nu}-\nabla_\mu \nabla_\nu f'
  +g_{\mu\nu}\Box f'=\kappa^2 T_{g\, \mu\nu}\,,
\end{equation}
where the prime stands for derivative with respect to $R$. The
trace reads
\begin{equation}\label{tracef}
  3\Box f'+f' R-2f=\kappa^2 T_g\,.
\end{equation}
In the spatially flat FRW's metric
\begin{equation}\label{FRWmetric}
 ds^2=dt^2-a^2(t)[dx^2+dy^2+dz^2]\,,
\end{equation}
Eqs. (\ref{fieldeqs}) and (\ref{tracef}) become
\begin{eqnarray}
-3\frac{\ddot a}{a}f'-\frac{f}{2}+3\frac{\dot a}{a}f'' {\dot
R}=\kappa^2 \rho\,, \label{0-0} \\
\left(\frac{\ddot a}{a}+2\frac{{\dot
a}^2}{a^2}\right)f'+\frac{f}{2}-2\frac{\dot a}{a}f''{\dot
R}-f'''{\dot R}^2-f'' {\ddot R}=\kappa^2 p\,, \label{i-j} \\
3f''' {\dot R}^2+3f'' {\ddot R}+9\frac{\dot a}{a}f'' {\dot
R}+f'R-2f=\kappa^2 T_g\, \label{trace}
\end{eqnarray}
where the dot denotes the derivative with respect to the cosmic
time, $\rho=T_{g\,0}^0$, $p\delta^i_j=T^i_{g\, j}$, and $T_g$ is
the trace of the energy-momentum tensor, $T_g=\rho-3p$. Moreover,
the Bianchi identities give a further condition on the
conservation of the energy
\begin{equation}\label{EnCons}
  {\dot \rho}+3\frac{\dot a}{a}(\rho +p)=0\,.
\end{equation}
To derive a solution to Eqs. (\ref{0-0})-(\ref{EnCons}), an
explicit form for $f(R)$ has to be specified. Following the models
proposed in literature, we make the ansatz
 \begin{equation}\label{f(R)}
 f(R)=\left(\frac{R}{A}\right)^n\,,
 \end{equation}
where $A$ is a constant with dimensions $m_P^{2-2/n}$ ($m_{P}\sim
1.22\,\, 10^{19}GeV$ is the Planck mass). By using Eq.
(\ref{f(R)}) and the ansatz $a(t)\sim t^{\alpha}$, Eqs.
(\ref{0-0}), (\ref{i-j}) and (\ref{EnCons}) imply
\begin{equation}\label{n=2alpha}
  n=2\alpha\,.
\end{equation}
One can easily check that the trace equation (\ref{trace}) is
fulfilled.

\begin{figure}
\centering \leavevmode \epsfxsize=7cm \epsfysize=5cm
\epsffile{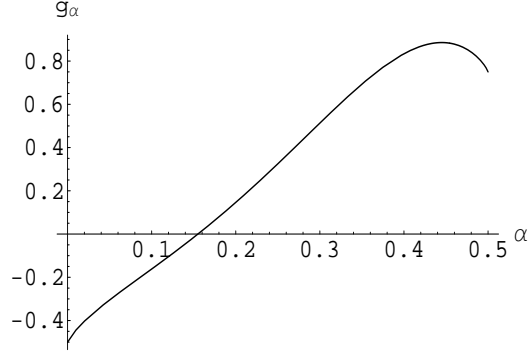} \caption{Plot of the function
$g_\alpha$ vs $\alpha$. The range in which the function $g_\alpha$
is positive, hence $T$ is defined, is $0.155\lesssim \alpha \leq
1/2$.} \label{nBsgAlphaLessHalf}
\end{figure}

The interaction (\ref{riccicoupling}) generates a net baryon
asymmetry provided ${\dot R}\neq 0$. As already noted, in standard
cosmology ${\dot R}$ vanishes during the radiation era, deviations
from the standard General Relativity prevent the Ricci curvature,
as well as its first time derivative, to vanish. Directly from the
definition of Ricci scalar curvature
 \[
 R=-6\left(\frac{\ddot a}{a}+\frac{{\dot a}^2}{a^2}\right)
 \]
it follows
\begin{equation}
 {\dot R}=-\frac{12\alpha(1-2\alpha)}{t^3}\,.
\end{equation}
The net baryon asymmetry (\ref{nB/s}) turns out to be
 \begin{equation}\label{basymt-T}
 \frac{n_B}{s}\simeq
 \frac{g_b}{g_*}\frac{45}{\pi^2}\frac{\alpha(1-2\alpha)}{t_D^3T_DM_*}\,,
\end{equation}
where $t_D$ is the decoupling time. Equating the expression of
$\rho$ given by Eq. (\ref{0-0}) to the usual expression of the
energy density \cite{kolb}
 \begin{equation}\label{rho(T)}
 \rho=\frac{\pi^2}{30}\, g_* T^4\,,
 \end{equation}
one obtains
\begin{equation}\label{T-t}
  T=\left(\frac{15}{4\pi^3 g_*}\right)^{1/4}
  g_\alpha^{1/4}\frac{m_P^{1/2}}{t^\alpha A^{\alpha/2}}\,,
\end{equation}
where
 \[
 g_\alpha\equiv
 6^{2\alpha}\alpha^{2\alpha}\frac{-10\alpha^2+8\alpha-1}{2(1-2\alpha)^{1-2\alpha}}\,.
 \]
Fig. \ref{nBsgAlphaLessHalf} shows $g_\alpha$ vs $\alpha$. The
allowed range for the parameter $\alpha$ is $0.155 \lesssim \alpha
\leq 1/2$ (the value $\alpha=1/2$ corresponds to Einstein's theory
of gravity). Nevertheless, as we shall discuss below, BBN imposes
stringent restrictions on its possible values

Inserting (\ref{T-t}) into (\ref{basymt-T}), one has
 \begin{equation}\label{basymfin}
 \frac{n_B}{s}\lesssim 5.8\, 10^{-3}H_\alpha\left(\frac{A}{m_P^{2-\frac{1}{\alpha}}}\right)^{\frac{3}{2}}
 \left(\frac{T_D}{m_{Pl}}\right)^{\frac{3}{\alpha}-1}
 \left(\frac{m_{Pl}}{M_*}\right)^2
\end{equation}
where
 \[
H_\alpha\equiv \frac{1}{\sqrt{\alpha|1-2\alpha|}}\left[
 876,5\frac{2(1-2\alpha)}{-10\alpha^2+8\alpha-1}\right]^{3/4\alpha}\,.
 \]
As pointed out in \cite{steinhardt}, a possible choice of the
cutoff scale $M_*$ is $M_*=m_{Pl}/\sqrt{8\pi}$ if $T_D=M_I$, where
$M_I\sim 2 \,\,10^{16}GeV$ is the upper bound on the tensor mode
fluctuation constraints in inflationary scale \cite{riotto}. This
choice is particular interesting because implies that tensor mode
fluctuations should be observed in the next generation of
experiments. For $A\sim m_P^{2-1/\alpha}$, using the constraint on
the observed baryon asymmetry $n_B/s\lesssim 9\,\, 10^{-11}$, Eq.
(\ref{basymfin}) can be recast in the form
\begin{equation}\label{BAymFin}
  \Lambda_\alpha \lesssim 1.6 \,,
\end{equation}
where
\begin{equation}\label{LambdaDef}
  \Lambda_\alpha\equiv 10^{12}H_\alpha (2.75\,\,
  10^{-3})^{3/\alpha}\,.
\end{equation}

Eq. (\ref{BAymFin}) is the main result of our paper: It relates
the estimation of the baryon number asymmetry inferred from
observations to the parameter $\alpha$, hence to the exponent of
the Ricci scalar curvature entering in the generic function
$f(R)$. This allows to fix the form of the $f(R)$-Lagrangian.

Let us now discuss $f(R)$-theories of gravity in the context of
BBN. According to BBN, the formation of light elements in the
early time occurred when the temperature of the Universe was
$T\lesssim 100$ MeV and the energy and the number density were
dominated by relativistic particles. At this stage of the
evolution of the Universe, the smattering of neutrons and protons
does not contribute in a relevant way to the total energy density.
All these particles are in thermal equilibrium owing to their
rapid collisions. Besides, protons and neutrons are kept in
thermal equilibrium by their interactions with leptons
\begin{eqnarray}
 \nu_e+n &\,\, \longleftrightarrow \,\, & p+e^- \label{int1} \\
 e^++n &\,\, \longleftrightarrow \,\, & p+{\bar \nu}_e \label{int2} \\
 n &\,\, \longleftrightarrow \,\, & p+e^- +{\bar \nu}_e \label{int3}
\end{eqnarray}
The weak interaction rate $\Lambda(T)$ is determined by means of
the conversion rates of protons into neutrons, $\lambda_{pn}(T)$,
and the inverse ones $\lambda_{np}(T)$. The rate $\lambda_{np}(T)$
is expressed in terms of the sum of rates associated to the
individual processes (\ref{int1})-(\ref{int3})
\cite{bernstein,kolb}
\begin{equation}\label{rate}
  \lambda_{np}=\lambda_{n+\nu_e \rightarrow  p+e^-}+
  \lambda_{n+e^+ \rightarrow  p+{\bar \nu}_e}+
  \lambda_{n \rightarrow  p+e^- +{\bar \nu}_e}\,.
\end{equation}
Similarly for $\lambda_{pn}(T)$. At enough high temperatures
($T\gg Q$ and $T\gg m_e$, where $Q=m_n-m_p\simeq 1.293MeV$, and
$m_{n,p,e}$ stands for the mass of the neutron, proton and
electron, respectively) the weak interaction rate is given by
\cite{kolb,bernstein}
\begin{equation}\label{Lambda(T)1}
\Lambda(T)=\lambda_{np}(T)+\lambda_{pn}(T)\simeq \frac{7\pi
(1+3g_A^2)}{60}G_F^2 T^5\,,
\end{equation}
where $G_F\simeq 1.166\,\, 10^{-5}GeV^{-2}$ is the Fermi coupling
constant and $g_A\simeq 1.26$ is the axial-vector coupling
constant of the nucleon. To compute the freeze-out temperature
$T_f$ (the temperature at which baryons decouple from leptons),
one has to equate $\Lambda(T)$ given by (\ref{Lambda(T)1}) to the
expansion rate of the Universe $H={\dot a}/{a}$: $\Lambda(T)
\simeq H$. Using (\ref{T-t}) with $g_*$ replaced by
$g_*^{(BBN)}=10.75$, it follows
\begin{equation}\label{Tfreeze}
  T_f= \left[6.4 \,\,
  10^{28-\frac{19}{\alpha}}\frac{\alpha}{g_\alpha^{1/4}}\left(\frac{4\pi^3g_*^{(BBN)}}{15}
  \right)^{\frac{1}{4\alpha}}
  \right]^{\frac{\alpha}{5\alpha-1}}GeV\,.
\end{equation}

\begin{figure}
\centering \leavevmode \epsfxsize=7cm \epsfysize=5cm
\epsffile{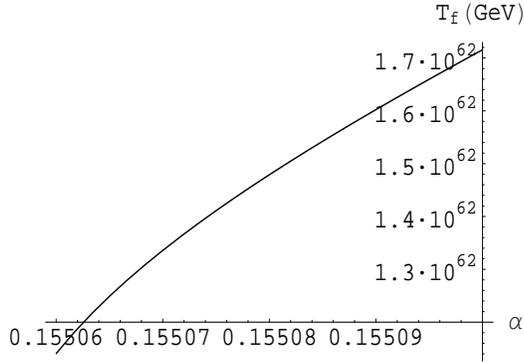} \caption{$T_f$ vs $\alpha$. In this plot the
parameter $\alpha$ assumes values $\gtrsim 0.15506$} \label{Tf0}
\end{figure}

\begin{figure}
\centering \leavevmode \epsfxsize=7cm \epsfysize=5cm
\epsffile{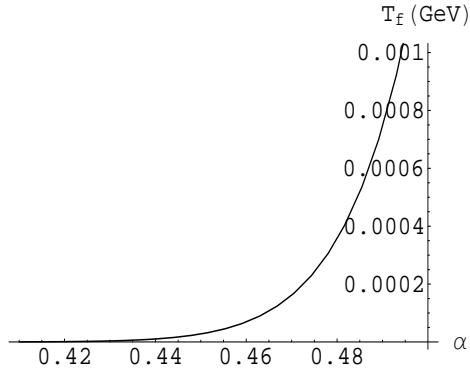} \caption{$T_f$ vs $\alpha$. In this plot the
parameter $\alpha$ assumes values $\leq 0.5$.} \label{Tf}
\end{figure}

\begin{figure}
\centering \leavevmode \epsfxsize=7cm \epsfysize=5cm
\epsffile{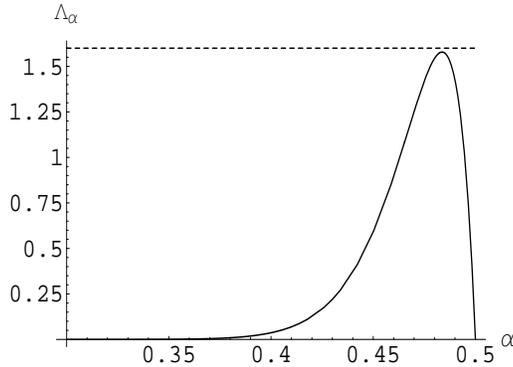} \caption{Plot of $\Lambda_\alpha$ vs
$\alpha$. The dashed line represents the upper bound $1.6$. The
maximum value $\sim 1.59$, corresponding to $\alpha\simeq 0.4845$,
agrees with the estimate (\ref{BAymFin}). Moreover, in the limit
$\alpha\to 1/2$, $\Lambda_\alpha$ vanishes, so no net baryon
asymmetry is generated (in agreement with General Relativity).}
\label{Lambda}
\end{figure}

Coming back to Eqs. (\ref{BAymFin}) and (\ref{LambdaDef}), the
analysis of the gravitational baryogenesis indicates that the
value $\alpha\simeq 0.15506$ is compatible with the estimation
(\ref{BAymFin}). Nevertheless, such a result is inconsistent with
BBN. As matter of fact, the plot of $T_f$ vs $\alpha$ in Fig.
\ref{Tf0} shows that values of $\alpha$ close to $0.155$ have to
be discarded because they lead to temperatures incompatible with
the typical temperatures of BBN ($(0.1-100)MeV$). This is not the
case for those values of $\alpha$ close to $0.5$, as arises from
Fig. \ref{Tf}. Thus we shall be interested only in the
gravitational baryogenesys for $\alpha \lesssim 0.5$. The results
are reported in Fig. \ref{Lambda}. As we can see, a narrow region
for the parameter $\alpha$, $0.46 \lesssim \alpha \lesssim 0.49$,
is consistent with the order of magnitude given in
(\ref{BAymFin}). In particular, the maximum value of the function
$\Lambda_\alpha$, $\Lambda_\alpha^{(max)}\sim 1.59$, is in
excellent agreement with the estimate (\ref{BAymFin}).
$\Lambda_\alpha^{(max)}$ corresponds to $\alpha\simeq 0.4845$ so
that $n\simeq 0.97$. The gravity Lagrangian therefore reads ${\cal
L}\sim R^{0.97}$, i.e. the form of the function $f(R)$ is {\it
nearly} linear in $R$.
We also note that in the limit $\alpha\to 1/2$, the function
$\Lambda_\alpha$ (hence $H_\alpha$) vanishes and no net baryon
asymmetry is generated. This is as expected because $\alpha=1/2$
implies that $a(t)$ evolves as $t^{1/2}$, i.e. the conformal
symmetry is restored, and therefore the Universe expands according
to General Relativity field equations.

In conclusion, $f(R)$-theories of gravity provide a natural arena
in which the baryon asymmetry in the Universe may be generated
through the mechanism of the gravitational baryogenesis. As we
have shown, deviations from General Relativity, although very tiny
(a few percent), allow to account for the observational data on
the matter-antimatter asymmetry. These results rely on the exact
solution of $f(R)$-field equations (Eqs. (\ref{0-0}) and
(\ref{i-j})), which allows to determine (without a fine-tuning)
the exponent $n$ for the generic function $f(R)\sim R^n$. The
results are also consistent with BBN.

\vspace{0.3in}

\acknowledgments The authors thank A. Iorio for comments and
reading the paper, and aknowledge support for this work provided
by MIUR through PRIN Astroparticle Physics 2004, and by research
funds of the University of Salerno.


\begin{thebibliography}{99}
\bibitem{noteJ} $J^\mu$ may be replaced by any current which yields a net $B-L$ charge in
thermal equilibrium; here $B$ and $L$ stand for baryon and lepton
number, respectively.
\end{thebibliography}
\end{document}